\documentclass[prb,twocolumn,amsmath,amssymb,a4page]{revtex4}

\usepackage{graphicx}
\usepackage{pifont}
\usepackage{color}

\begin{document}

\title{Piezoresistive heat engine and refrigerator}

\author{P.G. Steeneken, K. Le Phan, M.J. Goossens, G.E.J. Koops, G.J.A.M. Brom, C. van der Avoort and J.T.M. van Beek}
\affiliation{NXP-TSMC Research Center, NXP Semiconductors, HTC 4, 5656 AE Eindhoven, the Netherlands.}

\begin{abstract}
Heat engines provide most of our mechanical power and are essential for transportation on macroscopic scale. However, although significant progress has been made in the miniaturization of electrostatic engines, it has proven difficult to reduce the size of liquid or gas driven heat engines below 10$^7\mu$m$^3$.

Here we demonstrate that a crystalline silicon structure operates as a cyclic piezoresistive heat engine when it is driven by a sufficiently high DC current. A 0.34~$\mu$m$^3$ engine beam draws heat from the DC current using the piezoresistive effect and converts it into mechanical work by expansion and contraction at different temperatures. This mechanical power drives a silicon resonator of 1.1$\times$10$^3$~$\mu$m$^3$ into sustained oscillation. 
Even below the oscillation threshold the engine beam continues to amplify the resonator's Brownian motion. When its thermodynamic cycle is inverted, the structure is shown to reduce these thermal fluctuations, therefore operating as a refrigerator.
\end{abstract}

\maketitle

Most heat engines operate by the cyclic expansion and contraction of a gas at different temperatures. Much effort has gone into miniaturizing these conventional heat engine concepts\cite{Epstein04,Jacobson03, Spadaccini08}. However this downscaling is difficult, because the efficiency and power density of these engines reduce strongly with its size\cite{Peterson98}. A promising route for the miniaturization of heat engines is the use of an electrically heated solid as a working substance, because the large heat capacitance of solids allows much higher power densities and electrical interconnect facilitates energy transport and localized heat generation at the microscale. This concept has often been applied in electrothermal actuators for microscopic devices\cite{Wilfinger68, Elwenspoek89,Lammerink91,Guckel92,Reichenbach06, Seo08}, however the similarity between thermal expansion actuators and heat engines has received little attention. 

In order to operate as a cyclic heat engine, a thermal actuator needs to be supplemented with a mechanism that regulates the thermodynamic cycle of heating, expansion, cooling and compression. This cycle can be regulated by an external AC frequency synthesizer\cite{Lammerink91} or by a feedback loop, where a sustained reciprocating motion is generated from a DC power\cite{Wilfinger68,Reichenbach06,Seo08}. In previous reports of electrothermal engines, the thermodynamic cycle was always regulated by external electronic transistor circuits and amplifiers.

Here it is shown that a well-dimensioned silicon crystal of uniform composition operates as a cyclic heat engine when driven by a DC current. It is demonstrated that a silicon engine beam can drive a silicon resonator into sustained oscillation. In contrast to earlier work\cite{Wilfinger68,Reichenbach06,Seo08}, no external transistors or amplifiers are needed to regulate the heat engine, because a mechanism based on the intrinsic piezoresistive heating and thermal expansion of silicon provides the required feedback. The engine displacement volume of the heat engine, which similar to car engines is defined by the space occupied by the working substance, is 0.34~$\mu$m$^3$. Despite its low efficiency, the engine's power density is almost a factor 1000 higher than that of modern car engines. 

\begin{figure*}[t]
\includegraphics[scale=1]{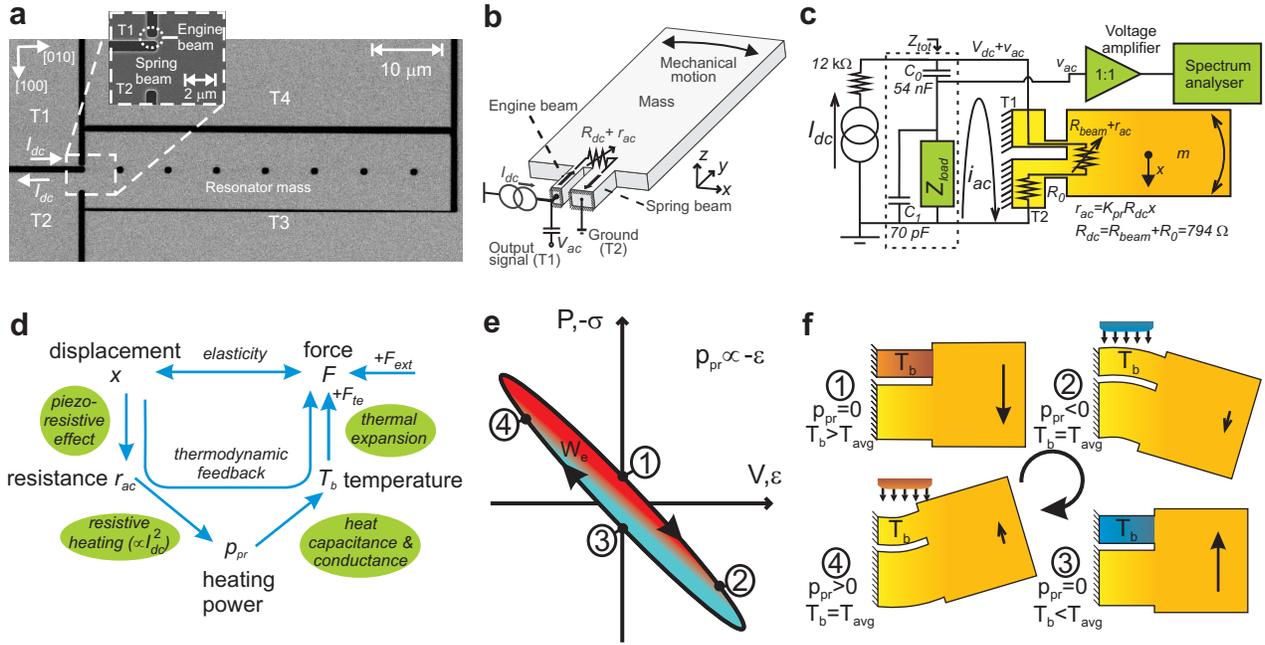}
\caption{\label{fig1} \textbf{Micrograph, schematics (not to scale), measurement circuit and operation mechanism of the piezoresistive heat engine and refrigerator.} \textbf{a} Scanning electron microscope (SEM) image of the heat engine. The inset shows a magnification of the wide spring beam and narrow engine beam by which the resonator mass $m$ is suspended. \textbf{b} Schematic of the device and electrical connections used for the measurements in Fig. \ref{fig2}b,c,d. \textbf{c} Schematic of the device and electrical circuit used for the measurements in Fig. \ref{fig3} and \ref{fig4}.  \textbf{d} Besides the elastic coupling between displacement and force, a thermally delayed feedback mechanism via the internal electrical and thermal variables occurs in a piezoresistive spring in the presence of a DC current flow. \textbf{e} Schematic thermodynamic cycle in the piezoresistive spring, when the device is operating at a single frequency as a heat engine. As a result of thermal delay, the temperature $T_b$ and stress $\sigma$ are phase shifted with respect to the strain $\varepsilon$ and piezoresistive heating power $p_{pr}$. The axes indicate the longitudinal strain $\varepsilon$ and stress $\sigma$, which are proportional to the volume change $V$ and pressure $P$ respectively. Therefore the engine beam expands at a higher temperature than at which it is compressed. It thus generates an amount of mechanical work $W_{e}$ from piezoresistive heat during each cycle. \textbf{f} Illustration of the phases \ding{192}-\ding{195} of the thermodynamic cycle shown in \ref{fig1}e, which are identified by the AC resistive heating power $p_{pr}$ and beam temperature $T_b$. To illustrate the position dependent heating power a fictitious external heat sink (blue) and source (red) are drawn. In reality the internal heating power in the beam depends on its position as a result of the piezoresistive effect.}
\end{figure*}

\begin{figure*}[t]
\includegraphics[scale=0.9]{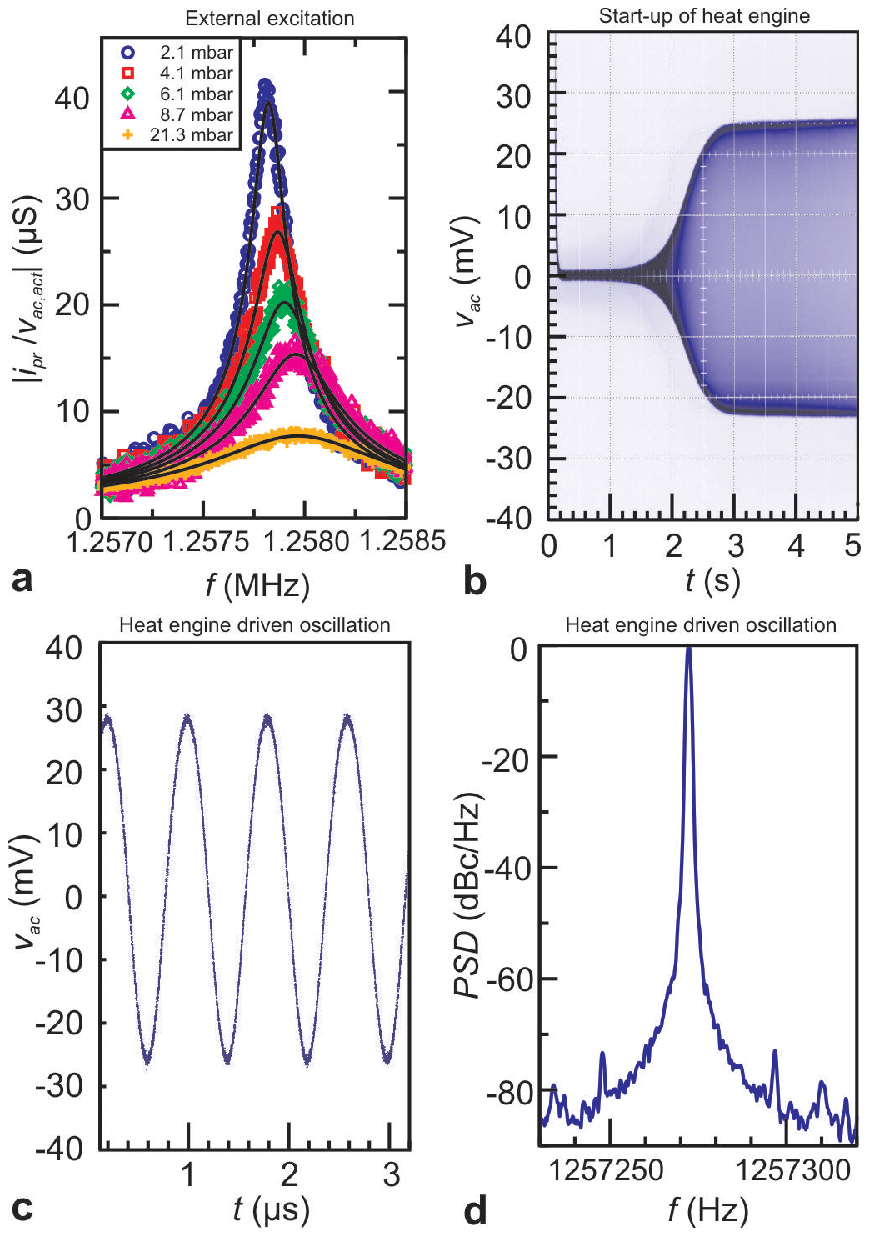}
\caption{\label{fig2} \textbf{Heat engine output signal.} \textbf{a} Characterization of the in-plane mechanical bending resonance by external excitation at different chamber pressures. \textbf{b,c,d} Operation of the heat engine at $I_{dc}$=$I_{t}$+0.01 mA at a chamber pressure $P$=0.01 mbar. In panel \textbf{b} the spontaneous start-up of the engine is measured by an analogue oscilloscope when a current $I_{dc}$=1.20 mA is switched on at $t$=0 s. The self-sustained sinusoidal output signal is measured \textbf{c} by a digital oscilloscope and \textbf{d} by a spectrum analyser that determines the power spectral density (PSD). From the oscilloscope data in panel \textbf{c}, the amplitude $x_0$ of the centre-of-mass is estimated to be $x_0$=43 nm, using $x_0=v_{ac,0}/(V_{dc}K_{pr})$, with $V_{dc}$=0.99 V and an AC amplitude $v_{ac,0}$=27 mV.}
\end{figure*}

Like any other reciprocating heat engine, this piezoresistive heat engine requires a minimum threshold power $P_t$ to compensate its internal dissipation and sustain its cyclic motion. In the first part of this article the operation of the piezoresistive heat engine above this threshold power $P_t$ is demonstrated by driving a resonator into sustained oscillation. In the second part the motion of the device at driving powers below $P_t$ is investigated. Measurements show that the engine beam continues to increase the thermal energy stored in the resonance mode like a heat pump. When the thermodynamic cycle of the engine is reversed, it is shown to reduce the stored thermal energy in the resonant mode, similar to recently developed optical and RF refrigerators\cite{Heidman99, Metzger04, Heidman06, Kleckner06, Schliesser06, brown07NIST, Metzger08, Lehnert09}. In contrast to conventional refrigerators, which simultaneously remove thermal energy from many degrees of freedom of a substance, this type of refrigerators cools by removing thermal energy from only a single mechanical resonance mode. This piezoresistive refrigerator provides an alternative way to reduce undesirable Brownian fluctuations in micromechanical sensors and mirrors.

\section*{Piezoresistive heat engine}
Experiments are performed on the homogeneous crystalline silicon structure shown in Fig. \ref{fig1}a. The structure consists of a mass measuring 12.5$\times$60.0$\times$1.5 $\mu$m$^3$, which is suspended by a 3 $\mu$m wide spring beam and a 280 nm narrow engine beam. Both beams have a length of 800 nm. Before discussing the stand-alone operation of the heat engine, its in-plane mechanical resonance at 1.26 MHz (Fig. \ref{fig2}a) is characterized by actuating it via terminal T3 with an AC electrostatic force and detecting the strain variations in the engine beam using the piezoresistive effect (see Methods). To operate the device as a piezoresistive heat engine, the external AC actuation voltage is disconnected from terminal T3. Thus, as shown in Fig. \ref{fig1}b the device is only connected at terminal T1 to a DC current source $I_{dc}$ and to a capacitively coupled oscilloscope. All other terminals are grounded. At low values of $I_{dc}$ only noise is observed on the oscilloscope. However, if the DC current is increased above a threshold $I_{t}$=1.19 mA a remarkable effect occurs: the device spontaneously starts to oscillate and generates a sinusoidal output voltage $v_{ac}$ with a frequency of 1.26 MHz. Oscilloscope and spectrum analyser measurements of the signal are shown in Fig. \ref{fig2}b, \ref{fig2}c and \ref{fig2}d. The motion of the device is also studied by stroboscopic illumination under a microscope. The stroboscope is triggered by the electrical output $v_{ac}$ of the heat engine. By adjusting the phase delay of the stroboscope a slow-motion movie of the engine's sinusoidal motion is made (see Supplementary Video). Stroboscopic and electrical measurements show that just above $I_{t}$ the displacement amplitude of the mass depends very sensitively on the DC current $I_{dc}$. If $I_{dc}$ is increased more than 50 $\mu$A above $I_{t}$, the optically observed mechanical amplitude $x$ stops increasing and is saturated by collisions of the resonator mass against electrode T3. This amplitude saturation is also observed in the piezoresistance variations $r_{ac}$.

The observed oscillation of the heat engine is attributed to the thermodynamic feedback mechanism which is schematically shown in Fig. \ref{fig1}d. When an alternating mechanical displacement $x=x_0 e^{i \omega t}$ with frequency $\omega$ is applied to a piezoresistive beam with spring constant $k$, this will not just result in an elastic force, but will also result in a resistance variation $r_{ac}$ via the piezoresistive effect. If a constant DC current $I_{dc}$ is flowing through the spring, the resistive heating power therefore changes by $p_{pr}=I_{dc}^2 r_{ac}$ and causes a temperature change, which in turn will generate a thermal expansion force $F_{te}$ that adds to the external forces $F_{ext}$ on the spring, such that $k x=F_{te}+F_{ext}$.  

The amplitude and the thermal phase-delay of the thermal expansion feedback force $F_{te}$ can be described by a complex coefficient $\beta$ which is defined by $F_{te}=\beta I_{dc}^2 k x$. The coefficient $\beta$ can be calculated from the material constants and geometry of the spring (see Supplementary Discussion A). The response of the piezoresistive spring to an external force can now be described by a complex spring constant $k^*_{\rm eff}\equiv F_{ext}/x=k(1-\beta I_{dc}^2)$. When the piezoresistive spring is coupled to a mass $m$, it forms a resonator that can store energy $U_i$ similar to the flywheel of a conventional heat engine. In the absence of current ($I_{dc}=0$ A), the resonator's intrinsic damping can be expressed by a $Q$-factor $Q_{\rm int}$ and its resonance frequency is given by $\omega_0=\sqrt{k/m}$. In the small-signal approximation, the linearised equation of motion of the heat engine driven resonator is equivalent to the well-known harmonic oscillator: $m\ddot{x}+\frac{\sqrt{k m}}{Q_{int}} \dot{x}+k^*_{\rm eff}x=0$. 
For a sinusoidal motion $x=x_0 e^{i \omega t}$ near the resonance frequency $\omega_0$ the imaginary (damping) part of this equation becomes zero at a threshold current value $I_{t}^2\!\! = \!\! (Q_{\rm int}{\rm Im}\beta)^{-1}$. Above this DC current $I_{dc}>I_t$ the power generated by the heat engine beam exceeds the intrinsic losses and the resonator will be brought into sustained oscillation by the engine beam.

The heat capacitance of the beam results in a thermal delay between piezoresistive heating power and temperature in the engine beam. As a consequence, the beam expands at a higher beam temperature $T_b$ than at which it contracts. The amplitude of temperature variations during the oscillation measured in Fig. \ref{fig2}c is estimated to be 0.2 K by finite element simulations. During each period of oscillation the silicon working substance in the piezoresistive beam therefore goes through a thermodynamic cycle, similar to a Stirling engine, during which it converts part of the resistive heat into work $W_{e}$. The piezoresistive beam therefore operates as a heat engine.  

Figure \ref{fig1}e shows the $P-V$ diagram and Fig. \ref{fig1}f the phases of the thermodynamic cycle of the engine. In phase \ding{192} the speed of the mass results in an expansion of the engine beam at high temperature. In phase \ding{193} the strain in the beam is tensile, such that it is cooling because the piezoresistance of $n$-type silicon is low. As a consequence the temperature of the beam is reduced in phase \ding{194} such that the beam contracts at low temperature, until it reaches phase \ding{195} where the compressive strain causes a high piezoresistive heating power, resulting again in an elevated temperature in phase \ding{192}. 

\begin{figure*}[t]
\includegraphics[scale=1]{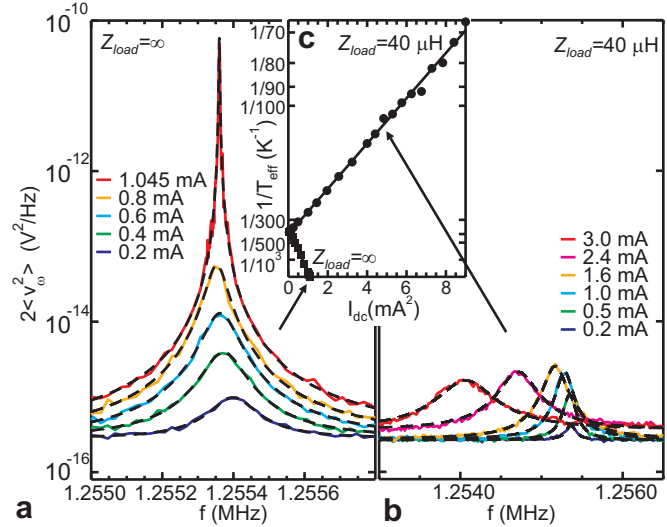}
\caption{\label{fig3} \textbf{Voltage spectral density $\! <\!\! v^2_{\omega}\!\! >$ for different values of $I_{dc}$.} Measurements by a spectrum analyser using the circuit in Fig. \ref{fig1}c show that \textbf{a} in the open state ($Z_{\rm load}=\infty$) the thermal motion is amplified when the heat engine is driven by a current $I_{dc}$. \textbf{b} For $Z_{\rm load}=40 \mu$H, the mechanical energy decreases with increasing current and the device operates as a refrigerator. Note the factor 4 difference in frequency scale. \textbf{c} $1/T_{\rm eff}$ as determined from the data in \ref{fig3}a and \ref{fig3}b by evaluating $\frac{1}{2}k_B T_{\rm eff}=\frac{1}{2}k<\!\! x^2\!\! >$, where $<\!\! x^2\!\! >$ is determined by integration of the fitted dashed lines (see Methods). The difference between the $y$-axis intersection of the data and the ambient temperature is within experimental uncertainty limits.}
\end{figure*}

\section*{Heat pump and refrigerator}

After having shown the operation of the heat engine above the oscillation threshold $I_{t}$, we now investigate its characteristics at lower DC currents. Below the threshold the displacement $x$ of the resonator will still exhibit random thermally excited 'Brownian' motion. The spectral displacement noise density $\! <\!\! x^2_\omega\!\! >\!$ of this motion near the resonance frequency in the presence of the feedback force $F_{te}$ can be determined from the fluctuation-dissipation theorem\cite{Heidman99,Kleckner06,Metzger08}:
\begin{eqnarray}
\label{x2}
<\!\! x^2_{\omega}\!\! > & = & \frac{2 \omega_0^3 k_B T_b/(k Q_{\rm int})}{(\omega^2-\omega_0^2)^2+(\omega \omega_0/Q_{\rm eff})^2} \\
\label{Qeff}
\frac{1}{Q_{\rm eff}} & = & \frac{1}{Q_{\rm int}}-I_{dc}^2{\rm Im}~\beta
\end{eqnarray}
Where $k_B$ is Boltzmann's constant and it is assumed that $Q_{\rm eff}\gg 1$ and $|\beta I_{dc}^2|\ll 1$.

To determine $<\!\! x^2_\omega\!\! >\!$, the AC voltage spectrum $<\!\! v^2_\omega\!\! >\!$, which is proportional to $<\!\! x^2_\omega\!\! >\!$, is measured (Fig. \ref{fig3}a,b) at several values of $I_{dc}$ by a spectrum analyser using the measurement circuit in Fig. \ref{fig1}c. In Fig. \ref{fig3}b an inductor $Z_{\rm load}=40 \mu$H is placed in parallel to the structure. Equation (\ref{x2}) is used to fit the data (see Methods).
The measurements in Fig. \ref{fig3}a show that even below the threshold current $I_t$ the thermal fluctuations in the resonance mode are amplified. It is observed that the stored thermal energy in the resonance mode increases with DC current, such that it equals the energy which would be stored in the mode if the resonator would be at an effective temperature $T_{\rm eff}$ which is defined by $\frac{1}{2}k_B T_{\rm eff}\equiv \frac{1}{2} k \int_{-\infty}^\infty \!\! <\!\! x^2_\omega\!\! >\! \frac{{\rm d}\omega}{2\pi}=\frac{1}{2}k <\!\! x^2\!\! >$. Similar to resonators cooled by optical or RF methods\cite{Heidman99, Metzger04, Heidman06, Kleckner06, Metzger08, Schliesser06, Lehnert09, brown07NIST} the temperature $T_{\rm eff}$ represents the mechanical energy stored in the resonance mode corresponding to the degree of freedom $x$ and is not representative for the other degrees of freedom of the resonator. By integrating the fitted $\! <\!\! x^2_\omega\!\! >\!$ curves,  $T_{\rm eff}$ is determined and is plotted in Fig. \ref{fig3}c. Depending on the load impedance $Z_{\rm load}$, the effective temperature $T_{\rm eff}$ of the resonance mode either increases like in a heat pump (Fig. \ref{fig3}a) or is refrigerated down to an effective temperature of 70 K at $I_{dc}$=3.0 mA (Fig. \ref{fig3}b).

The difference between the spectra in Figs. \ref{fig3}a and \ref{fig3}b is caused by the fact that the feedback coefficient $\beta$ is proportional to the ratio $\gamma_Z\equiv p_{pr}/(I_{dc}^2 r_{ac})$ between the piezoresistive heating power $p_{pr}$ and the AC resistance $r_{ac}$ as can be seen from Fig. \ref{fig1}d. Therefore the coefficient $\beta$ can be written as the product of two complex numbers: $\beta=\gamma_Z \beta_0$, where $\beta_0$ depends solely on the device geometry and material parameters and $\gamma_Z$ can be controlled externally via $Z_{\rm load}$ in the following way. Figure \ref{fig1}c shows that a resistance change $r_{ac}$ of the beam will induce both an AC current $i_{ac}=-v_{ac}/Z_{\rm tot}$ and AC voltage $v_{ac}=R_{\rm dc}i_{ac}+I_{dc}r_{ac}$. The total heating power $P_{dc}+p_{pr}$ in the engine beam is given by $(I_{dc}+i_{ac})^2(R_{beam}+r_{ac})$ and for $r_{ac}\ll R_{beam}$ it follows that $\gamma_Z=1-2 R_{beam}/(Z_{\rm tot}+R_{\rm dc})$. 

\begin{figure*}[t]
\includegraphics[scale=1]{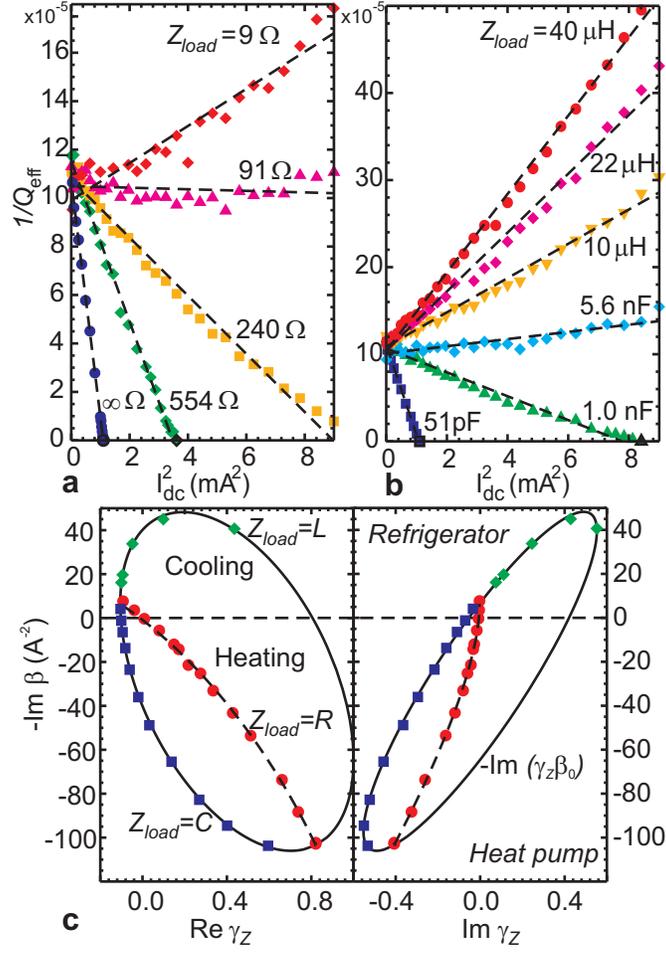}
\caption{\label{fig4} \textbf{Measurements and fits to confirm the operation mechanism proposed in Fig. \ref{fig1}d.} \textbf{a,b} Inverse effective $Q$-factor $1/Q_{\rm eff}$ versus DC current $I_{dc}^2$ for different values of the load impedances $Z_{\rm load}$ indicated in the graph. Dashed lines are fits using equation (\ref{Qeff}). Symbols on the $x$-axis correspond to the minimum current level at which self-sustained oscillation was observed. \textbf{a} Resistive load impedances $Z_{\rm load}$. \textbf{b} Capacitive and inductive load impedances $Z_{\rm load}$. \textbf{c} Measured slopes $-{\rm Im}~\beta$ of the curves in Fig. \textbf{a},\textbf{b} plotted along the real (left) and the imaginary (right) $\gamma_Z$-axis for resistive (red circle), capacitive (blue square) and inductive (green diamond) values of $Z_{\rm load}$. An excellent fit of the data is obtained by plotting $-{\rm Im}~(\gamma_Z \beta_0$) with $\beta_0$=$-123.7+64.6i A^{-2}$, for resistive (dashed lines), capacitive and inductive (solid lines) values of $Z_{\rm load}$. If ${\rm Im}~\beta$ is positive the engine beam acts as a heat pump that increases the effective temperature and if ${\rm Im}~\beta$ is negative the engine beam acts as a refrigerator that cools the effective temperature of the resonance mode.}
\end{figure*}

\section*{Quantitative analysis}
Voltage spectra like those in Fig. \ref{fig3}a,b are recorded for several real and imaginary values of $Z_{\rm load}$, using resistors, capacitors and inductors. The fitted values of $Q_{\rm eff}$ are plotted in Fig. \ref{fig4}a and b. For small values of $Z_{\rm load}$ a more accurate determination of $Q_{\rm eff}$ was made by generating a random electrostatic noise force on the resonator using a white voltage noise source connected to terminal T3. The observed linear dependence of $1/Q_{\rm eff}$ on $I^2_{dc}$ in Fig. \ref{fig4}a,b corresponds well with equation (\ref{Qeff}). From the slope of these linear fits, ${\rm Im}~\beta$ is determined and is plotted against $\gamma_Z$ as symbols in Fig. \ref{fig4}c. As shown by solid and dashed lines in Fig. \ref{fig4}c, an excellent multiple linear regression fit of the data is obtained using the function ${\rm Im}~\beta={\rm Im}~(\gamma_Z \beta_0)$, which yields the fit parameters $\beta_0=-123.7+64.6i$ A$^{-2}$ and $R_{beam}=439.3$ $\Omega$.  

In Supplementary Discussion A an analytical model is derived for the complex Young's modulus and spring constant $k^*_{\rm eff}$ and it is used to derive an estimate for $\beta_0$, which yields $\beta_0=-132+157i$ A$^{-2}$. A finite element simulation of the full geometry including the anisotropy of the silicon crystal results in $\beta_0=-111+58i$ A$^{-2}$. Both the excellent fits in Fig. \ref{fig3} and \ref{fig4}, and the quantitative agreement between the measured and simulated values of $\beta_0$ support the proposed feedback mechanism in Fig. \ref{fig1}d. 

\begin{figure*}[t]
\includegraphics[scale=1]{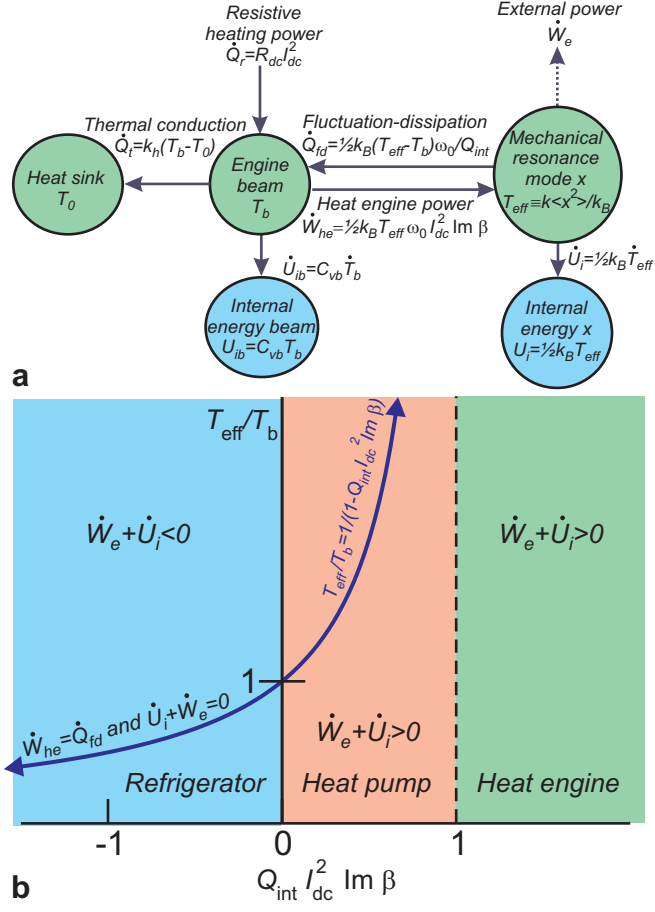}
\caption{\label{fig5} \textbf{Diagrams showing the heat flow and operation modes of the engine beam.} \textbf{a} Rate equations governing the energy transport between heat sink, engine beam and resonance mode. The resistive heating $\dot{Q}_{r}$ powers the engine beam, which converts the heat into work $\dot{W}_{he}$ on the resonance mode. The resonance mode dissipates the power $\dot{W}_{he}$ at a rate $\dot{Q}_{fd}$ and uses the excess power to perform external work $\dot{W}_{e}$ or to increase $T_{\rm eff}$. Heat from the engine beam flows away to the heat sink by its thermal conductance $k_h$. Internal energy ($U_{ib}$,$U_i$) is stored by the heat capacitances of the beam at constant volume ($C_{vb}$) and of the resonance mode ($\frac{1}{2} k_B$). All quantities in this diagram are time averaged values and the dots represent time derivatives. Integration of equation (\ref{x2}) yields\cite{Heidman99,Kleckner06,Metzger08} the relation $T_{\rm eff}/T_b=Q_{\rm eff}/Q_{\rm int}$ and combining this with equation (\ref{Qeff}) and the definition of $Q_{\rm int}$ results in the equations for $\dot{Q}_{fd}$ and $\dot{W}_{he}$. \textbf{b} Three different regimes of operation are distinguished. Depending on the feedback factor $Q_{\rm int}I_{dc}^2 {\rm Im} \beta$, the device operates as a refrigerator, heat pump or heat engine. If no external work or heat is transferred ($\dot{W}_e+\dot{U}_i$=0), the effective temperature of the resonance mode follows the blue line $T_{\rm eff}/T_b=(1-Q_{\rm int}I_{dc}^2 {\rm Im} \beta)^{-1}=Q_{\rm eff}/Q_{\rm int}$.}
\end{figure*}

\section*{Flow rates of heat and work}
In order to gain insight in the operation of the heat engine, a schematic showing the rate equations for the flow of heat and work between electrical source, heat sink, engine beam and resonance mode is shown in Fig. \ref{fig5}a. All rates shown in this figure are time averaged, such that all variations which cancel each other during a single cycle are averaged out. The three modes of operation which were experimentally observed in Fig. \ref{fig2}d, \ref{fig3}a, and \ref{fig3}b are shown in Fig. \ref{fig5}b. 
For Im $\beta <0$, $\dot{W_{he}}$ is negative, thus cooling the effective temperature of the resonance mode and operating as a refrigerator. For $0<Q_{\rm int} I_{dc}^2 {\rm Im} \beta<1$ heat is pumped into the resonance mode, increasing $T_{\rm eff}$ until a new steady state is reached at which $\dot{W}_{e}+\dot{U}_{i}=0$. In this mode the engine beam operates as a heat pump. For $Q_{\rm int} I_{dc}^2 {\rm Im} \beta>1$, the steady state rate equation in Fig. \ref{fig5}a can only be satisfied with $\dot{W}_{e}+\dot{U}_{i}>0$. Therefore external work $W_{e}$ is generated and the engine beam operates as a heat engine. The distinction between the work and heat flows in Fig. \ref{fig5} is subtle. For $T_{\rm eff}\rightarrow \infty$ it follows that $Q_{\rm eff}$ tends to infinity such that the motion of the resonator is perfectly sinusoidal, like in Fig. \ref{fig2}c. This energy can theoretically be stored with 100\% efficiency and should be called work. At any finite value of $T_{\rm eff}$ the resonator performs a thermally driven random motion, which cannot be fully converted into work according to Carnot's theorem\cite{Feynman63} and should be called heat.

\section*{Performance}

For the miniaturization of engines a high power density $p_{dens}$ is essential, since it determines the minimum size the engine can have in order to perform its task within the specified time. The maximum power $\dot{W}_{he}$ that the demonstrated heat engine can generate is given by $\frac{1}{2}k<x^2>\omega_0 I_{dc}^2 {\rm Im} \beta$. In order to achieve the maximum power density, a current $I_{dc}$=5 mA is applied for which the amplitude of the centre of mass is limited by collisions with electrode T3 to $x$=100 nm. For an optimum value of Im $\beta$=100 A$^{-2}$ (see Fig. \ref{fig4}c) it is found that $\dot{W}_{he}$=12.7 nW, about 4\% of which is dissipated by the intrinsic losses ($\dot{Q}_{fd}$). It thus follows that the piezoresistive heat engine, with an engine displacement volume of the engine beam of $V=0.34$~$\mu$m$^3$, has a power density $p_{\rm beam}=\dot{W_e}/V$=37 GW/m$^3$ which is almost a factor 1000 higher than that of car engines which have typical power densities of around 50 MW/m$^3$. The power density $p_{\rm beam}$ of the piezoresistive heat engine is also high when compared to other microengines \cite{Epstein04,Spadaccini02,Koser01}, which have reported power densities up to 3 GW/m$^3$. However, in some of these cases the power density is defined using the volume of the total structure instead of the engine displacement volume. Due to the relatively large size of the proof mass compared to the engine beam size, this definition of power density is less favourable for the piezoresistive heat engine, yielding  $p_{\rm tot}$=11 MW/m$^3$.

Using the equations from Fig. \ref{fig5}a, the efficiency of the presented engine is calculated  to be $\eta_{he}=(\dot{W}_{he}-\dot{Q}_{fd})/\dot{Q}_r=$6.4$\times 10^{-7}$. Besides being limited by the Carnot theorem, the efficiency of the piezoresistive heat engine is also limited by the fracture limit of silicon to around $\eta_{\rm pr,Si,max}=3.4\times 10^{-4}$, as analysed in the Supplementary Discussion B. It is therefore expected that the efficiency and power density of the piezoresistive heat engine can still be improved by a factor $\sim$500 by increasing the displacement amplitude and thermal design of the engine. Even after optimization, this efficiency is low compared to that of micro- and nano-engines driven by electrostatic forces\cite{Fennimore03,Fan05,Ayari07,Weldon10}. Nevertheless, the comparison in Supplementary Discussion B shows that piezoresistive heat engines have the potential to outperform electrostatic engines in terms of power density. This high power density and the absence of transistors can facilitate the development of mechanical actuator\cite{Stemme06} and sensor\cite{Seo08, Burg07} arrays with very high element densities.

Similar to other recently investigated cooling methods\cite{Heidman99, Metzger04, Heidman06, Kleckner06, Schliesser06, brown07NIST, Metzger08, Lehnert09}, the coefficient of performance of the refrigerator $\eta_{\rm cool}=-\dot{W}_{he}/\dot{Q}_r$ is very small when $T_{\rm eff}$ is at or below room temperature. The refrigerator will therefore not be useful for lowering the temperature of systems with many degrees-of-freedom. However, because the heat capacity $\frac{\partial U_i}{\partial T_{\rm eff}}$ of a single-degree-of-freedom resonance mode is only $\frac{1}{2}k_B$ and because the thermal insulation of the mode is high as a result of its high $Q_{\rm int}$ (see Fig. \ref{fig5}a), the piezoresistive refrigerator can still reduce $T_{\rm eff}$ by more than 200 K. The Brownian fluctuations associated with the energy $\frac{1}{2}k_B T_{\rm eff}$ often dominate the noise in mechanical sensors and mirrors and their reduction can thus increase the signal-to-noise ratio and might ultimately enable quantum-limited measurement and control of mechanical resonators. The presented piezoresistive refrigerator provides an alternative cooling method for reducing these Brownian fluctuations, which has the advantage that it does not require laser or RF sources and only requires a silicon chip with a relatively low DC power.

\section*{Methods}
\textbf{Structure}
The piezoresistive heat engine in Fig. \ref{fig1}a is made out of a 1.5 $\mu$m thick silicon layer on a silicon-on-insulator wafer, with a phosphor doping concentration $N_d$=4.5$\times$10$^{18}$ cm$^{-3}$ giving a specific resistivity of $\rho_{dc}$=10$^{-4} \Omega$m. This doping concentration results in a low temperature coefficient of resistivity\cite{Bullis68} of less than 0.1\% per K (see Supplementary Fig. C). The thin crystalline silicon layer is structured in a single mask step by a deep reactive ion etch and the buried SiO$_2$ layer below the mass and beams is removed in a hydrogen fluoride vapour etch. A DC current $I_{dc}$ is driven through the beams via terminals T1 and T2 as shown in Fig. \ref{fig1}b and c. As a result of the geometry of the structure, the current density, heating power density and mechanical stress are concentrated in the narrow engine beam, but the resonance frequency of the resonator is mainly determined by the spring beam and resonator mass. The in-plane mechanical bending resonance mode that determines the fundamental resonance frequency $f_0$=1.26 MHz has a spring constant $k$=256 N/m as determined from finite element method (FEM) simulations. Measurements are performed at 315 K in a vacuum chamber at a pressure below 10$^{-2}$ mbar. 

\textbf{External excitation}
The in-plane fundamental mechanical resonance of the structure at 1.26 MHz is characterized by actuating it via terminal T3 with an AC electrostatic force $F_{\rm ac,act}=\varepsilon_0 A V_{\rm dc,act}v_{\rm ac,act}/g^2$ generated by a voltage $V_{\rm dc,act}+v_{\rm ac,act}$ on terminal T3, with $V_{\rm dc,act}$= -1 V. The actuation gap $g$=200 nm has an actuation area $A=60\times 1.5$ $\mu$m$^2$. A DC current $I_{dc}=0.1$ mA is applied as in Fig. \ref{fig1}b. The displacement of the centre of mass $x$ is proportional to the measured piezoresistive AC current $i_{pr}$, which is shown in Fig. \ref{fig2}a. The reduction of the intrinsic $Q$-factor $Q_{\rm int}$ upon increasing the pressure in the vacuum chamber confirms the mechanical nature of the resonance. The solid fits in Fig. \ref{fig2}a allow us to determine the piezoresistive factor $K_{pr}$, which is defined by $r_{ac}/R_{dc}=K_{pr} x$, where $r_{ac}$ is the amplitude of the piezoresistance variation and $R_{dc}$ is the total DC resistance. From the maximum piezoresistive current $i_{pr,peak}$ the piezoresistive factor is determined to be $K_{pr}=-\frac{i_{pr,peak}}{I_{dc}}\frac{k}{F_{\rm ac,act}Q_{\rm int}}=-6.6\times 10^5$ m$^{-1}$.

\textbf{Spectrum analysis}
The spectra in Fig. \ref{fig3} and \ref{fig4} are measured using the electrical circuit in Fig. \ref{fig1}c. Because the displacement of the centre-of-mass $x_\omega \equiv x(\omega)$ causes a piezoresistive voltage $v_{\omega} \equiv v_{ac}=[Z_{\rm tot}^{-1}+R_{\rm dc}^{-1}]^{-1} I_{dc} K_{pr} x$ (Fig. \ref{fig1}c), the electrical spectrum $<\!\! v^2_{\omega}\!\! >$ measured by the spectrum analyser is proportional to equation (\ref{x2}). The voltage spectral density in Fig. \ref{fig3}a and b is fit by dashed curves which are given by equation (\ref{x2}) with $<\!\! v^2_{\omega}\!\! >\! =I_{dc}^2 K_{pr}^2\!\! <\!\! x^2_{\omega}\!\! >\!\! /[Z_{\rm tot}^{-1}+R_{\rm dc}^{-1}]^2$ and which include a fit parameter to account for the white Johnson noise background. The total AC impedance parallel to the resonator $Z_{\rm tot}=(i\omega C_0)^{-1}+(i\omega C_1+Z_{\rm load}^{-1})^{-1}$ can be controlled by adjusting $Z_{\rm load}$. The DC resistance $R_{\rm dc}=R_{\rm beam}+R_0$=794 $\Omega$, bias and cable capacitances $C_0=54$ nF and $C_1=70$ pF were measured using an impedance analyser at $1.26$ MHz. Terminals T3 and T4 are grounded. A 1 M$\Omega$ input impedance voltage buffer amplifier is used to minimize the effects of cables and spectrum analyser on the engine operation. 
In Fig. \ref{fig3}b DC resistive heating can lead to a temperature increase of the beam which opposes the refrigeration mechanism. It is estimated from the temperature dependence of the resistance and from FEM simulations that the maximum temperature of the engine beam is 370$\pm$20 K at $I_{dc}=3$ mA (see Supplementary Fig. C). This temperature increase of 17\% by DC resistive heating is relatively small compared to the factor 5 decrease in $T_{\rm eff}$ as a result of the reduction of $Q_{\rm eff}$.
We thank J.J.M. Ruigrok, C.S. Vaucher, K. Reimann, R. Woltjer and E.P.A.M. Bakkers for discussions and suggestions and thank J. v. Wingerden for his assistance with the SEM measurements. The authors declare that they have no competing financial interests. Author Contributions: K.L.P., P.G.S., J.T.M.v.B. and M.J.G. invented and designed the device. P.G.S., K.L.P., M.J.G. and C.v.d.A. performed the experiments. P.G.S. developed the theory, analysed the experiments and wrote the article. J.T.M.v.B., G.E.J.K. and G.J.A.M.V. developed the process technology and manufactured the device.Correspondence and requests for materials should be addressed to P.G.S. (peter.steeneken@nxp.com).

\clearpage
\onecolumngrid
\renewcommand{\appendixname}{SUPPLEMENTARY DISCUSSION}
\section*{\large Supplementary Information \\ Piezoresistive heat engine and refrigerator}
\begin{center}
\parbox{16cm}{\small In this Supplementary Information the thermodynamic coupling between stress and strain in piezoresistive materials carrying a current is discussed in more detail. In Supplementary Discussion A an analytical expression is derived for the effective complex Young's modulus $Y^*_{\rm eff}$. This expression is used to provide an analytical estimate for the feedback coefficient $\beta$, which expresses the effect of a DC current on the effective spring constant $k^*_{\rm eff}$ of a piezoresistive spring. The analytical result is compared to measurements presented in the main manuscript. In Supplementary Discussion B the optimal efficiency and power density of silicon heat engines is estimated. In Supplementary Figure C a measurement of the resistance of the device as a function of DC current and chuck temperature is used to estimate the temperature rise in the engine beam.}
\end{center}
\appendix
\section{Quantitative discussion of the feedback mechanism and coefficient $\beta$}
To evaluate the effect of a DC electric current on the mechanical properties of a piezoresistive solid, consider a solid through which a current density $J_{dc}$ is flowing in the $y$-direction. When a small uniaxial vibrational stress $\sigma_{ac}e^{i \omega t}$ with frequency $\omega$ is present along the $y$-direction the piezoresistive effect will induce an AC resistivity change $\rho_{ac}$:
\begin{equation}
\label{rho}
\rho_{ac} e^{i \omega t}=\rho_{dc}\pi_l \sigma_{ac} e^{i \omega t}
\end{equation}
Where $\rho_{dc}$ is the unstressed resistivity and $\pi_l$ is the longitudinal piezoresistive coefficient \cite{Smith54b}. It is assumed that $\rho_{ac}\ll \rho_{dc}$. This resistivity change modifies the AC resistive heating power density $p_{pr}$:
\begin{equation}
\label{pac}
p_{pr} e^{i \omega t}=\gamma_Z J_{dc}^2 \rho_{ac}e^{i \omega t}
\end{equation}
As discussed in the main manuscript, $\gamma_Z$ is a correction factor which is needed if a finite impedance to ground is present parallel to the piezoresistive resonator, such that AC currents $i_{ac}$ can also contribute to $p_{pr}$. From the heat equation $p=-k_h \nabla^2 T+c_p \rho_d \partial T/\partial t$ it follows that the heating power causes sinusoidal temperature fluctuations with amplitude $T_{ac}$:
\begin{equation}
\label{T}
T_{ac} e^{i \omega t}=\frac{p_{pr} e^{i \omega t}}{\gamma_h+i \omega c_p \rho_d}
\end{equation}
The specific heat capacity is given by $c_p$ and $\rho_d$ is the mass density. The effect of thermal heat conductivity $k_h$ is captured by the factor 
$\gamma_h$:
\begin{equation}
\label{gh}
\gamma_h=-k_h (\nabla^2 T_{ac})/T_{ac}
\end{equation}
The factor $\gamma_h$ depends on the resonator's geometry as will be discussed below. The temperature increase generates a thermal expansion stress:
\begin{equation}
\label{te}
\sigma_{ac,te}=\alpha_{te} Y T_{ac}
\end{equation}
$Y$ is Young's modulus and $\alpha_{te}$ is the thermal expansion coefficient. This thermal expansion stress adds to the externally applied stress $\sigma_{ac,ext}$, such that the total stress is given by $\sigma_{ac}=\sigma_{ac,ext}+\sigma_{ac,te}$, and the corresponding strain is given by $\varepsilon_{ac}=\sigma_{ac}/Y$. 
The stress-strain relation of the piezoresistive solid is therefore the same as that of a solid with an effective complex Young's modulus $Y^*_{\rm eff}=\sigma_{ac,ext}/\varepsilon_{ac}$. The variables, coupling mechanisms and multiplicative factors in equations (\ref{rho}-\ref{te}) have been schematically shown in figure A1 and using these equations the effective modulus can be expressed as:
\begin{equation}
\label{Yeff}
Y^*_{\rm eff}(\omega)\equiv \frac{\sigma_{ac}-\sigma_{ac,te}}{\varepsilon_{ac}} \approx Y \left(1-\gamma_Z \frac{ \rho_{dc} \pi_l  J_{dc}^2 \alpha_{te} Y   }{\gamma_h+i \omega \rho_d c_p}\right)
\end{equation}
This relation is only valid for small stress and strain, such that the linear approximation is valid and material parameters are constant. When all constants in equation (\ref{Yeff}) are real and positive, equation (\ref{Yeff}) is identical to the Young's modulus of the standard anelastic solid \cite{Nowick72b}, however if some of the constants are negative or have a non-zero imaginary part, it can lead to properties like mechanical self-amplification and negative creep. This modification of the effective Young's modulus can occur in any piezoresistive solid in the presence of electrical current. Besides its effect on the dynamics of mechanical structures it might thus also affect the propagation of acoustic waves in solids. 
\begin{figure*}
\includegraphics[scale=0.4]{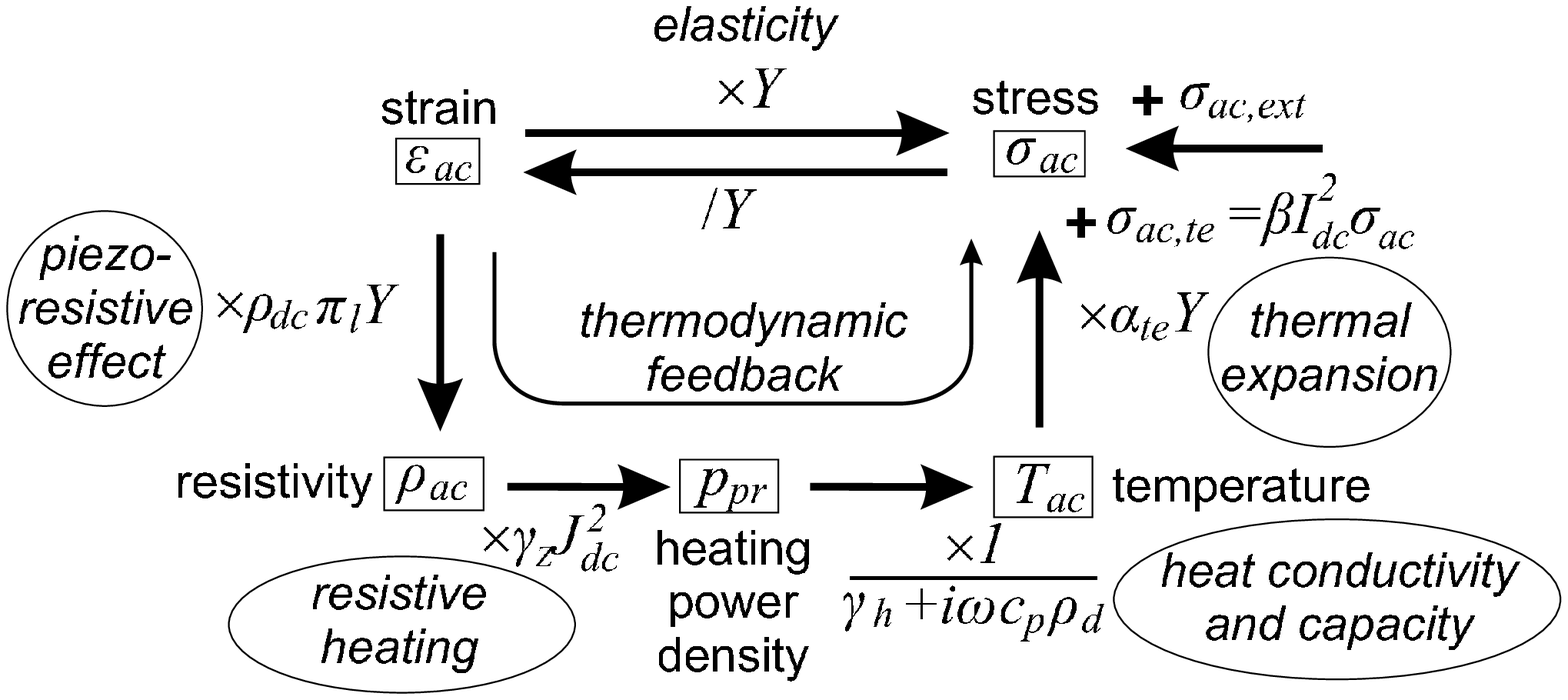}
\parbox{18cm}{\small \textbf{Supplementary Figure A:} Schematic of the thermodynamic coupling mechanism via the piezoresistive, resistive heating and thermal expansion effects. Relevant variables in the mechanical, electrical and thermal domain are shown. Coupling mechanisms are represented by arrows.}
\end{figure*}
As follows from appendix A of Nowick and Berry \cite{Nowick72b}, the effective spring constant of a resonance mode with a complex position dependent Young's modulus is given by:
\begin{equation}
\label{keff2}
k^*_{\rm eff}=k\left(\frac{\int_V Y^*_{{\rm eff}}({\bf z}) \varepsilon_{ac}^2({\bf z})d{\bf z}}{\int_V Y \varepsilon_{ac}^2({\bf z})d{\bf z}}\right)=k (1-\beta I^2_{dc})
\end{equation}
Where the integrals run over the volume $V$ of the resonator. Combining this with equation (\ref{Yeff}), it is found that the feedback coefficient $\beta$ can be expressed as:
\begin{equation}
\label{beta}
\beta=\gamma_k \gamma_Z \frac{ \rho_{dc} \pi_l \alpha_{te} Y   }{A^2 (\gamma_h+i \omega \rho_d c_p)}=\gamma_Z \beta_0
\end{equation}
Where $A$ is the cross-sectional area of the piezoresistive conductor, and $\gamma_k$ is a factor which accounts for the non-uniformity of $Y^*_{\rm eff}$. If $Y^*_{\rm eff}$ is constant over the whole volume, $\gamma_k=1$.

For the resonator under consideration, the thermodynamic feedback effect is large inside the narrow engine beam where the strain and current density are concentrated. It is therefore a good approximation to assume that $Y^*_{\rm eff}$ is given by equation (\ref{Yeff}) inside the volume $V_{\rm engine}$ of the engine beam and is equal to $Y$ everywhere else in the resonator. In this case the factor $\gamma_k$ is given by:
\begin{equation}
\label{gk}
\gamma_k=\frac{\int_{V_{\rm engine}} \varepsilon_{ac}^2({\bf z})d{\bf z}}{\int_{V} \varepsilon_{ac}^2({\bf z})d{\bf z}}
\end{equation}
For the lowest frequency in-plane bending resonance mode this fraction is found to be $\gamma_k=0.11$ using a finite element simulation.

To evaluate $\gamma_h$ the AC heat equation is to be solved:
\begin{equation}
\label{he}
k_h \nabla^2 T_{ac}+p_{pr}=i c_p \rho_d \omega T_{ac}
\end{equation}
Since the dimensions of the engine beam are much smaller than the thermal wavelength $\lambda_h=\sqrt{8\pi^2 k_h/(c_p \rho_d \omega_0)}\approx$ 26 $\mu$m, the heat equation can be simplified by assuming that the AC resistive heating density $p_{pr}$ is zero outside the engine beam and that $p_{pr}$ and $T_{ac}$ are independent of position inside the engine beam. The beam can be treated as a point source at $y=0$ with temperature $T_{ac,0}e^{i \omega t}$. The corresponding solution of the 1-dimensional heat equation outside the beam is:
\begin{equation}
\label{Txt}
T(x,t)=T_{ac,0}e^{-(1+i)2\pi |y|/\lambda_h +i\omega t}
\end{equation}
Since the heat conduction from the 2 ends of the beam needs to be equal to the difference between the generated and stored energy in the beam, it follows from equation (\ref{T}) that:
\begin{equation}
-2 k_h \nabla T_{ac}|_{y=0}=L (p_{pr}-i c_p \rho_d \omega T_{ac})=L \gamma_h T_{ac}
\end{equation}
Where $L$=800 nm is the beam length. Substitution of equation (\ref{Txt}) yields:

\begin{equation}
\label{gh2}
\gamma_h=4\pi (1+i) k_h/(L \lambda_h)
\end{equation}

The material constants along the [100] direction of $n$-type silicon are $Y=130$ GPa, $\pi_l=-102\times$10$^{-11}$ Pa, $\alpha_{te}=2.6\times 10^{-6}$ K$^{-1}$, $\rho_d=$2329 kg/m$^3$, $c_p=702$ J/kg$\cdot$K and $k_h=$113 W/(K$\cdot$m). The cross-sectional area is $A$=280$\times$1500 nm$^2$. Substituting these values in equations (\ref{gh2}) and (\ref{beta}) and using $\gamma_k$=0.11 from equation (\ref{gk}) yields $\beta_0=-132+157i$ A$^{-2}$. This value is higher than the experimental value, which is mainly attributed to the fact that the actual heat flux outside the beam is not 1-dimensional but radial, which results in a larger value of $\gamma_h$. A finite element simulation of the full geometry, including the anisotropy of the silicon crystal results\cite{Arxivb} in $\beta_0=-111+58i$ A$^{-2}$, in good agreement with $\beta_0=-123.7+64.6i$ A$^{-2}$ as determined from the fits of the measurements in figure 4 of the main manuscript.

\section{Efficiency and power density of silicon heat engines}

The optimal efficiency of any heat engine $\eta_{he,max}$ is limited by the Carnot efficiency $\eta_C=1-T_{\rm min}/T_{\rm max}$. In solid heat engines there is another efficiency constraint related to the maximum strain before fracture of the solid working substance. To illustrate this, consider a silicon beam with cross section $A$ and length $L$ going through an idealized Stirling cycle of heating at constant volume, expansion at constant high temperature $T_{\rm max}=T_{\rm min}+\Delta T$, cooling at constant volume and contraction at constant low temperature $T_{\rm min}$. The beam's spring constant is $k=Y A/L$ and the static thermal expansion is $\Delta x_{te}=\alpha_{te} \Delta T L$. The heat needed to raise the temperature of the beam is $Q_h=A L c_v \rho_d \Delta T$, where $c_v$ is the heat capacity at constant volume. The isothermal expansion over a distance $\Delta x$ cost external work $W_{\rm heat,e}=-\frac{1}{2} k (\Delta x-\Delta x_{te})^2$. To keep the temperature stable during expansion, additional heat $\sim Q_h (c_p-c_v)/c_v$ is needed. Then the beam is cooled at constant volume to temperature $T_{\rm min}$. During isothermal contraction, the beam generates external work $W_{\rm cool,e}=\frac{1}{2} k (\Delta x)^2$. The total external work done by the engine beam is therefore $W_e=W_{\rm heat,e}+W_{\rm cool,e}\approx k \Delta x \Delta x_{te}=\alpha_{te}Y A \Delta x \Delta T$, for $\Delta x_{te}\ll \Delta x$.  The fracture strain limit of silicon which is in the range\cite{Wilson96b} of  $\varepsilon_{\rm Si,frac}\approx$ 0.01-0.03, as a conservative estimate we use $\varepsilon_{Si,frac}=$ 0.01. The maximum expansion $\Delta x$ from compressive to tensile strain is given by be $\Delta x_{\rm max}=2 \varepsilon_{\rm Si,frac} L$. Therefore the optimal efficiency of a silicon thermal engine is estimated to be:
\begin{equation}
\label{etasimax}
\eta_{\rm Si,max}=W_{e}/Q_h\approx 2 \frac{Y \alpha_{te} \varepsilon_{\rm Si,frac}}{\rho_d c_p}=4 \times 10^{-3} 
\end{equation}
Higher efficiencies are possible if the strain is always kept compressive to enable higher $\Delta x$, or by using a regenerator to reuse the heat lost during cooling. Note that for quasi-static thermal actuators $\Delta x=\Delta x_{te}$ and the work done is $W_e=\frac{1}{2}k (\Delta x_{te})^2$, such that the optimal efficiency is lowered by a factor $\alpha_{te} \Delta T/(4 \varepsilon_{\rm Si,frac})$. Even for $\Delta T$=100 K, a quasi-static actuator will have an optimal efficiency that is a factor 150 lower than the optimal dynamic effiency. For most quasi-static actuators the efficiency will be much lower because the thermal conduction rate $\dot{Q}_t$ will not only cool during the isothermal contraction, but during the whole cycle.

Now we continue to estimate the maximum efficiency of the piezoresistive heat engine. From figure 5a in the main manuscript it is found that the maximum rate of work done by the heat engine is given by $\dot{W}_{he}=\frac{1}{2}k <x^2> \omega_0 I_{dc}^2 {\rm  Im} \beta$. The value of $\beta$ in equation (\ref{beta}) is maximal for n-type silicon ($\pi_l<0$) with $\gamma_k=1$, $\gamma_Z=1$. Optimally $\gamma_h$ is zero, however this would mean that there is no thermal conduction and no cooling of the heat engine. As a compromise $\gamma_h=\omega \rho_d c_p$ is taken, reducing the efficiency by a factor 2. Thus it is found from equation (\ref{beta}) that the maximum value of ${\rm Im} \beta$ for the presented piezoresistive heat engine is given by:
\begin{equation}
\label{betamax}
{\rm Im} \: \beta_{\rm pr,max}\approx -\frac{\rho_{dc} \pi_l \alpha_{te} Y}{2 A^2 \omega \rho_d c_p}=7500 \: {\rm A}^{-2}
\end{equation}

The highest measured value of Im $\beta$ is a factor 75 lower, of which a factor 9 can be attributed to $\gamma_k=0.11$, and the remaining factor $\sim$8 is mainly due to a too high value of $\gamma_h$. The heating power in the beam is given by $\dot{Q}_r=I_{dc}^2 \rho_{dc} L/A$. The spring constant is given by $k=Y A/L$ and the average amplitude $<x^2>=\frac{1}{2} (\varepsilon_{\rm Si,frac} L)^2$. Thus the optimal efficiency of the piezoresistive heat engine is found to be:
\begin{equation}
\label{etaprmax}
\eta_{\rm pr,Si,max}=\frac{\dot{W}_{he}}{\dot{Q}_r}\approx -\frac{\pi_l Y^2 \alpha_{te} \varepsilon_{\rm Si,frac}^2}{8\rho_d c_p}=3.4\times 10^{-4}
\end{equation}

A small value value of $\gamma_k$ does not reduce the efficiency, because the reduction in $\beta$ is compensated by the higher value of stored energy $\frac{1}{2}k <x^2>$ in the equation for $\dot{W}_{he}$. In other words, the efficiency of the heat engine beam is independent of the resonator to which it is coupled. For the heat engine reported in the main manuscript, the maximum strain in the engine beam is estimated with finite element simulation to be $\varepsilon_{b}\approx 0.0015$ when operating at an amplitude of the center of mass $x=$100 nm. Moreover the measured Im $\beta \gamma_k$ is a factor 8 lower than optimal, such that the estimated efficiency is $\eta_{\rm he,est}= \eta_{\rm pr,Si,max}\times (0.0015/0.01)^2/8=10^{-6}$ close to the measured value. The difference between the measured efficiency and equation (\ref{etaprmax}) suggest that there is still room for efficiency improvement by a factor 500 of the presented heat engine concept by increasing its amplitude and reducing the heat conductivity parameter $\gamma_h$. The equations for maximum efficiency \eqref{etaprmax} and \eqref{etasimax}, which are based on linear approximations, cannot exceed the Carnot efficiency and should obey $\eta_C>\eta_{\rm Si,max}>\eta_{\rm pr,Si,max}$.

Based on the heat equation of a 1D beam with fixed temperature at its ends, the DC temperature rise in the engine beam is estimated as $\Delta T_{\rm max}=\frac{\dot{Q}_r}{V}\frac{L^2}{8 k_h}$. The maximum power density of the piezoresistive heat engine is given by: 
\begin{equation}
\label{pdensmax}
p_{\rm pr,dens,max}=\frac{\dot{W}_{\rm he,max}}{V}=\eta_{\rm pr,Si,max}\frac{\dot{Q}_r}{V}\approx -\frac{\pi_l Y^2 \alpha_{te} \varepsilon_{\rm Si,frac}^2}{\rho_d c_p}\frac{k_h \Delta T_{\rm max}}{L^2} \approx 100 \frac{\rm TW}{{\rm m}^3}
\end{equation}
Where the equation has been evaluated at L=800 nm and $\Delta T_{\rm max}=200$ K. Even higher power densities can be achieved by reducing the beam length $L$. When decreasing $L$ the frequency $\omega$ needs to go up to maintain a sufficient efficiency according to the condition $\gamma_h \approx \omega \rho_d c_p$ and equation (\ref{gh2}). A good heat sink is needed to keep the ends of the beam at low temperature. 

Recently much progress has been made on engines operated by electrostatic forces\cite{Fennimore03b,Fan05b,Ayari07b,Weldon10b}. Although these engines can have a high efficiency, their work per cycle is limited by the maximum energy\cite{Baglio07b} that can be stored in the electric field ($E_{\rm e,max} \approx$4$\times 10^5$ J/m$^3$). Assuming a 100\% efficiency their power density is approximately limited by $p_{\rm el,dens,max}=E_{\rm e,max}\times f$. Comparing this with equation (\ref{pdensmax}) it is estimated that the power density of piezoresistive engines with a length $L=800$ nm will exceed that of electrostatic engines up to frequencies of 200 MHz. By lowering the length $L$ this limit can be pushed to even higher frequencies. Despite their low efficiency, it is thus anticipated that in most cases piezoresistive heat engines will outperform electrostatic engines in terms of power density. 

  \renewcommand{\appendixname}{SUPPLEMENTARY FIGURE}
    \renewcommand{\figurename}{\bf Supplementary Figure C\!\!}
    
\section{Temperature and current dependence of resistance}

\begin{figure}[h]
\includegraphics[scale=0.38]{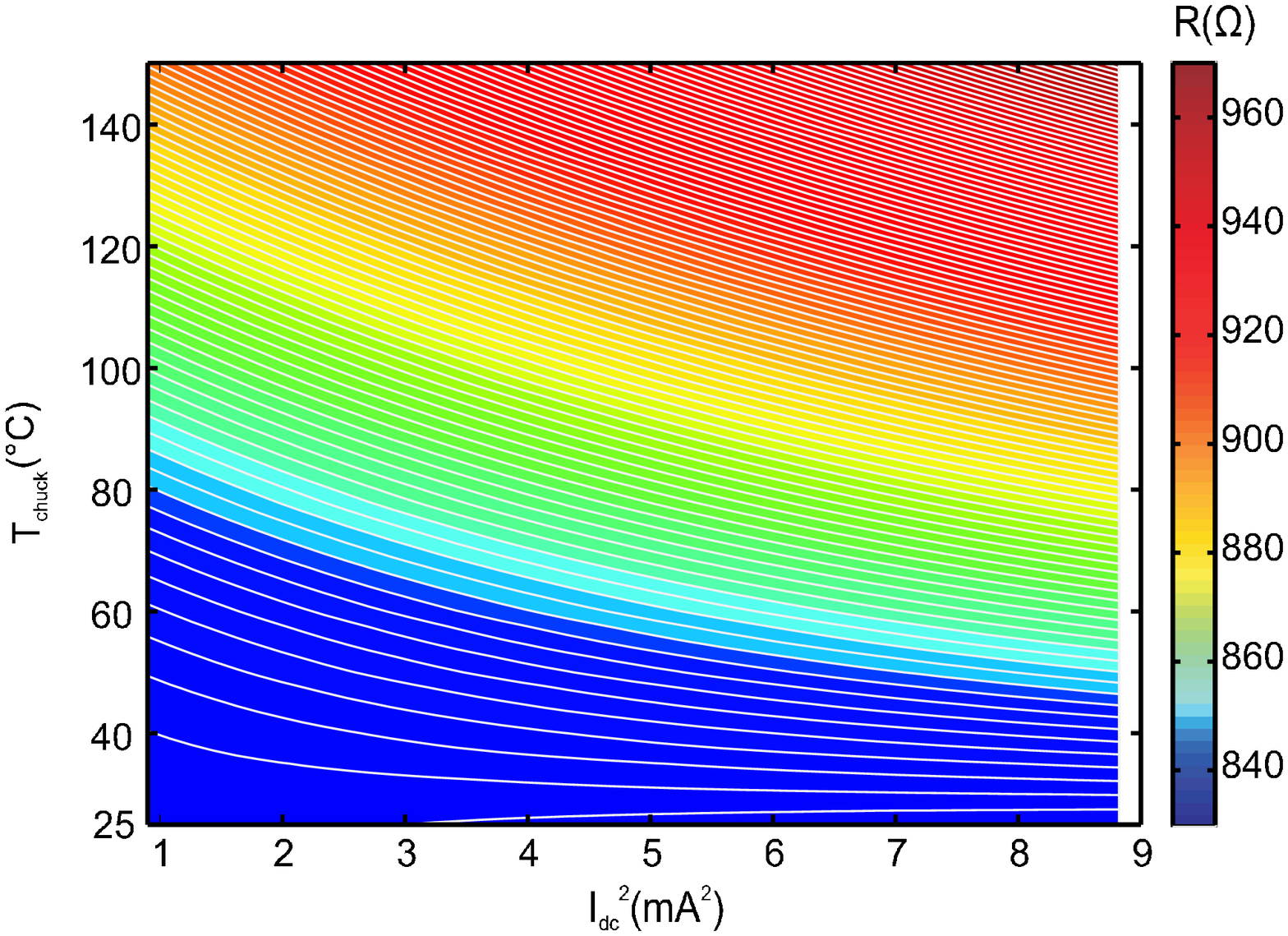}
\end{figure}
{\small \noindent {\bf Supplementary Figure C:} Measurement of the resistance of the resonator $R_{dc}$ as a function of $I_{dc}$ and chuck temperature $T_{\rm chuck}$. The difference in resistance between two equiresistance lines is 1.4 $\Omega$. For a uniformly heated device, the temperature increase $\Delta T$ is proportional to the DC power $R_{dc} I_{dc}^2$ and the slope of the equiresistance lines in the figure is expected to be constant. In the measured temperature range, the slope of the equiresistance lines is between $\Delta T/I_{dc}^2$=0-8~K/(mA$^2$). A FEM simulation of the structure with a constant (temperature independent) resistivity of $\rho_{dc}$=10$^{-4} \Omega$m predicts $\Delta T/I_{dc}^2$=3.75~K/(mA$^2$) for the maximum engine beam temperature\cite{Arxivb}. The observed variation in the slope is due to the non-uniform temperature distribution in the structure in combination with the strong temperature dependence of TCR, which is near a minimum and is even observed to be slightly negative at room temperature\cite{Bullis68b}. As a consequence of this, the total resistance of the structure is almost independent of $I_{dc}$ around $T_{\rm chuck}$=30$^\circ$C, where the increase of resistance of the warmer parts compensates the decrease of resistance of the colder parts of the structure. From these data it is estimated that in practice $\Delta T/I_{dc}^2$=6$\pm$2~K/(mA$^2$), such that the maximum temperature rise in the beam is estimated to be $\Delta T$=56$\pm$18 K at 3 mA at maximum cooling power and $\Delta T$=7$\pm$2 K when operating as a heat engine at 1.05 mA.}

\section*{Supplementary References}

\end{document}